# A Bioplausible Model for the Expanding Hole Illusion: Insights into Retinal Processing and Illusory Motion

Nasim Nematzadeh[1], David M. W. Powers[1,2]
College of Science and Engineering [(1)] & Caring Futures Institute [(2)]
Flinders University of South Australia
Adelaide, Australia

nasim.nematzadeh, david.powers@flinders.edu.au

## Abstract

The *Expanding Hole Illusion* is a compelling visual phenomenon in which a static, concentric pattern evokes a strong perception of continuous forward motion. Despite its simplicity, this illusion challenges our understanding of how the brain processes visual information, particularly motion derived from static cues. While the neural basis of this illusion has remained elusive, recent psychophysical studies [1] reveal that this illusion induces not only a perceptual effect but also physiological responses, such as pupil dilation. This paper presents a computational model based on Difference of Gaussians (DoG) filtering and a classical receptive field (CRF) implementation to simulate early retinal processing and to explain the underlying mechanisms of this illusion. Based on our results we hypothesize that the illusion arises from contrast-dependent lateral inhibition in early visual processing. Our results demonstrate that contrast gradients and multi-layered spatial processing contribute to the perception of expansion, aligning closely with psychophysical findings and supporting the role of retinal ganglion cells in generating this illusory motion signal. Our findings provide insights into the perceptual biases driving dynamic illusions and offer a new framework for studying complex visual phenomena.

Keywords— Visual perception; Bioinspired neural networks; Retinal Ganglion Cells; Classical Receptive Field (CRF); Difference of Gaussian (DoG); Motion illusions; Expanding Hole Illusion

## 1. Introduction

Visual illusions provide a valuable insight into the functioning of the human visual system, as they reveal discrepancies between physical reality and perceptual interpretation. Such illusions demonstrate how the brain relies on specific assumptions and processing mechanisms to construct our experience of the world [2]. One particularly intriguing example is the *Expanding Hole Illusion*, which creates a sense of motion in a static visual pattern. This illusion features a central dark region surrounded by concentric, high-contrast gradients that give the impression of dynamic expansion despite remaining stationary.

Two prominent theories that attempt to explain similar phenomena include Hering's theory of assimilation and contrast [3] and Changizi's theory of perceiving present from the past [4]. Each offers insights into the perceptual mechanisms underlying illusions, yet they operate on distinct principles that are also relevant to the *Expanding Hole Illusion*.

Hering's theory of assimilation and contrast [3] suggests that the visual system enhances or diminishes local contrasts based on surrounding elements, leading to perceived motion or depth effects. This theory is often applied to illusions where adjacent colours or patterns influence each other, creating an exaggerated response in perceived contrast or movement. According to Hering, the visual system groups similar elements while distinguishing contrasting ones, thereby producing the illusion of motion or expansion. In the *Expanding Hole Illusion*, the gradual radial gradient in luminance towards the darker central region may evoke depth or movement through contrast effects.

Changizi's theory [4] of perceiving present from the past, in contrast, postulates that certain illusions result from the brain's attempt to predict the immediate future states based on previous sensory input, compensating for the neural processing delay. This theory posits that the brain anticipates changes in a scene to account for the time it takes for visual information to be processed. In the *Expanding Hole Illusion*, the darkening centre, surrounded by gradient shading, could be interpreted as movement towards the observer. Changizi's theory [4] would suggest that the illusion stems from a predictive response, where the brain interprets the gradient pattern as forward motion in anticipation of where the observer would be if the image were actually dynamic.

While previous research has primarily focused on its effects on visual perception, Laeng et al. [1] went a step further by demonstrating a compelling physiological response: the illusion causes the pupil to dilate, evoking a sensation of a gradual loss of light, like entering a dark tunnel. They also noted: "In contrast, central light patterns of different colours (e.g., green, magenta, and also white) will suggest an expansion of light, as when moving toward a differently illuminated space" [1]. This finding suggests that the visual system perceives the expanding pattern as real motion, prompting an adaptation in the eye's optical system. Although related illusions, such as the Café Wall [5] and Tilt illusions [6], have been explored extensively, the underlying mechanisms of the *Expanding Hole Illusion* remain less understood.

In Fig 1., we illustrate two variations of the *Expanding Hole Illusion*, which demonstrate the impact of contrast polarity on visual perception. The panel on the left shows the original *Expanding Hole Illusion*, where a static pattern with a central dark region and concentric gradients creates the impression of forward movement, as if the observer were approaching a tunnel or hole. This illusion generates a compelling sense of optic flow, mimicking the experience of motion despite the image's static nature. The central black "hole" surrounded by lighter gradients appears to expand outward, likely triggering neural pathways associated with motion detection and visual depth.

On the right, we present an inverted version of the illusion, known as the "*White Hole*" variation. Here, the colours are reversed, with white ellipses replacing the dark central region against a black background. This contrast polarity reversal leads to a different physiological response, with many observers reporting pupil constriction, mimicking the sensation of moving toward a bright light. This effect underscores the role of retinal and cortical processing in interpreting motion and brightness changes based on contrast cues, contributing to our understanding of how visual illusions engage neural mechanisms for motion and depth perception.

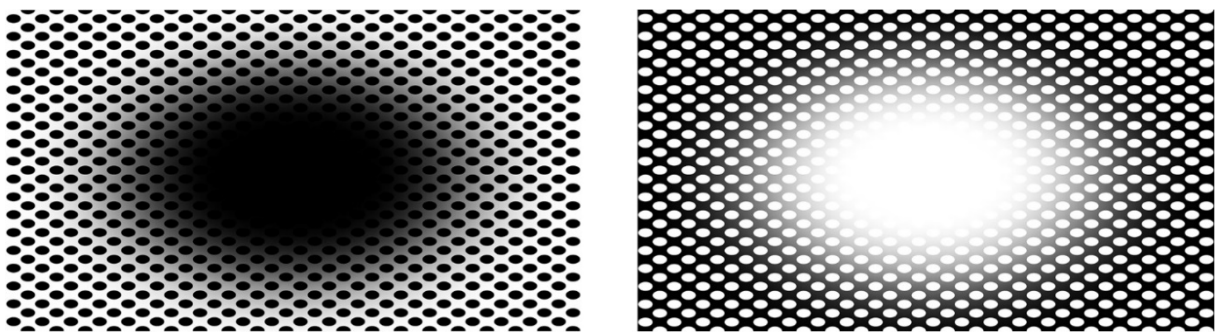

*Fig. 1. (Left) The Expanding Hole Illusion: this static pattern creates a dynamic effect, giving the impression of continuous expansion of the central region in the pattern. It creates the impression of forward movement into a dark tunnel. (Right) The contrast polarity reversal of the pattern on the left referred to as "White Hole" version, where the observers' pupils contracted instead of expanding, as if moving toward a bright light.*

Despite the psychophysical evidence linking this illusion to the visual system, the specific underlying mechanisms responsible have not been thoroughly investigated. In this paper, a model is proposed based on the classical Difference of Gaussians (DoG) filter, which replicates the receptive field properties of retinal ganglion cells (RGC). This model aims to elucidate how the visual system could misinterpret static stimuli as dynamic, offering a framework to understand how contrast processing constructs the illusion of expansion.

Thus, while Hering's and Changizi's theories [3,4] emphasise cognitive and interpretive processes, the retinal ganglion cell-based model offers a biological explanation anchored in the retinal structure, contributing to our understanding of how visual illusions can originate from fundamental neural circuitry.

## 2. Related Work

The study of visual illusions has long been a key avenue for understanding the complexities of the human visual system. Illusions highlight the brain's reliance on assumptions and specific processing mechanisms, exposing discrepancies between physical stimuli and their perceptual interpretations [7]. Research has shown that low-level and high-level visual processes play crucial roles in generating these effects [2,8]. Various motion illusions, such as the Rotating Snakes and Fraser-Wilcox illusions, have been explored using models based on gradient-based processing and motion detectors, which emphasize the importance of contrast-sensitive mechanisms [9-11]. These models primarily rely on changes in luminance and perceptual grouping to simulate the perception of illusory motion. Specifically, research by Whitney & Cavanagh [11] demonstrated the importance of perceptual grouping in motion perception, highlighting how the organization of visual elements can influence the perception of motion in various illusions. However, the effectiveness of these approaches in explaining complex patterns like the *Expanding Hole Illusion* remains limited. Several models have been proposed to explain such illusions, with classical approaches relying on contrast and orientation selectivity in early visual areas [12]. Building upon this, our model extends these principles to account for additional features like the perceived expansion of static images.

Recent advancements in understanding the *Expanding Hole Illusion*, specifically through the work of Laeng et al. [1], have demonstrated a coupling between perceptual effects and physiological responses. Laeng et al. found that the illusion induces pupil dilation, suggesting that the visual system interprets the static pattern as indicative of real motion. This finding aligns with Kitaoka's [13] earlier work on dynamic illusions, which proposed that these phenomena can lead to physiological responses resembling those triggered by actual movement. The expanding pattern's ability to evoke such responses implies that early visual areas may interpret the illusion as a real-world motion stimulus, adjusting the optical system accordingly. This research supports the idea that the visual system interprets the static pattern as representing real depth or movement, causing the brain to adjust the pupil size accordingly.

Moreover, Larsen et al. [14] explored neural adaptation and its role in perceiving static motion illusions, contributing critical findings that complement our research. This work offers a valuable foundation for understanding how neural dynamics shape the perception of both motion and geometric illusions. Our model builds upon these findings, offering quantifiable predictions based on differences in neural activations induced by illusory stimuli.

In addition, Kitaoka's work [15] focused extensively on various types of illusions, including the *Expanding Hole Illusion*, which has gained widespread recognition. While their research primarily concentrated on the psychophysical effects, our approach diverges by emphasizing computational modelling to predict and analyse the underlying neural activities.

In this context, we draw on the established properties of retinal ganglion cells, which are known to exhibit center-surround receptive fields. This feature allows them to respond differentially to local changes in luminance and contrast [16]. The work of Hubel & Wiesel [8] further elucidated the organization of visual pathways and the processing of visual information in the retina and cortex, providing a foundation for understanding how visual illusions can be generated from underlying neural mechanisms.

Previous research has demonstrated the efficacy of neurophysiological models [17] in predicting geometric visual illusions, with a particular focus on the role of neural responses in the perception of these phenomena. One such model employed Difference of Gaussians filtering, a classical approach to simulating receptive fields of retinal ganglion cells, demonstrating how contrast-dependent lateral inhibition can contribute to illusions like the Café Wall and Tilt illusions [5,6,18]. By incorporating multi-layered spatial and contrast-sensitive filters, the model accounts for the underlying mechanism for perceiving the illusory cues in these patterns. Other studies have supported this concept, emphasising the role of contrast gradients and differential neural activation in the perception of various dynamic illusions [19,20].

In addition, Pinna and Brelstaff [21] introduced a novel visual illusion involving relative motion, further emphasizing the importance of contrast and spatial relationships in generating illusory movement. Their work demonstrated that even minor modifications in visual stimuli can elicit substantial shifts in perceived motion, reinforcing the notion that illusions arise from early neural interactions sensitive to contrast changes. Similarly, Conway et al. [9] demonstrated the critical role of receptive fields in motion perception, illustrating how localized contrast variations activate neural circuits that drive the sensation of movement.

In summary, the existing body of research underscores the significance of contrast-sensitive mechanisms, perceptual grouping, and neural interactions in generating motion illusions. Laeng et al.'s [1] findings on physiological responses to the *Expanding Hole Illusion*, combined with the established role of retinal processing [16] and contrast-based filtering models, set the stage for further exploration of how static patterns can evoke such compelling illusions of motion.

Our approach to explaining the *Expanding Hole Illusion* focuses on retinal ganglion cell (RGC) responses and uses a model that mimics classical receptive fields (CRFs) to implement lateral inhibition effects in the retina. This approach grounds the explanation in low-level, retinal processing, where ganglion cells are known to enhance or suppress signals based on surrounding patterns. In this model, the centre-surround organization of retinal ganglion cells selectively responds to the luminance gradient of the illusion pattern. The central darkening, with its surrounding lighter gradients, creates strong lateral inhibition at the boundary, which leads to an exaggerated perception of motion as the retina differentially responds to contrast changes within the pattern. This explanation not only provides a mechanistic basis for understanding the illusion but also aligns with how early visual processing in the retina contributes to perceived motion and depth even from static images.

Here, by leveraging DoG filtering, we aim to replicate these neural mechanisms and provide a computational explanation for the *Expanding Hole Illusion*. Our proposed model builds on established neurophysiological principles and incorporates contrast-based filtering to account for the perceived expansion.

# 3. Methodology

Recent physiological findings have advanced our understanding of retinal ganglion cells (RGCs) and their functional diversity. Studies have shown that both the retina and the mammalian visual cortex employ multiscale representation and processing strategies, supported by physiological and psychophysical evidence [8,22,23]. In a comprehensive study on retinal circuitry and visual coding, Field and Chichilnisky [22] identified at least 17 distinct types of RGCs in the retina, each with a specialised role in encoding visual information . The characteristics and size of receptive fields vary with eccentricity (distance from the fovea) and retinal circuitry, providing evidence for multiscale encoding mechanisms within the retina. While it was once thought that orientation selectivity was confined to the cortex, it has been found that certain retinal cells also exhibit orientation selectivity similar to cortical cells, aligning with Marr's theory of vision and its raw-to-full primal sketch concept [24,25].

The centre-surround organisation of RGC receptive fields is commonly attributed to lateral inhibition (LI) within both the outer and inner retinal layers [26]. At the first synaptic level, this LI mechanism [27] enhances photoreceptor signals by enabling activated cells to inhibit nearby cells, creating a retinal pulse response or point spread function (PSF). This biological convolution effect serves to enhance edges [28] in visual perception, functioning as a bandpass filter that supports essential visual tasks. Inhibition occurring at the second synaptic level in the inner retina is believed to contribute to more complex response properties, such as directional selectivity [26]. In our model, we utilise the contrast sensitivity of RGCs, structured around a circular centre-surround organisation for the retinal receptive fields [29,30].

### 3.1. DoG Filtering for Simulating Retinal Processing

The Difference of Gaussians (DoG) filter is a widely used computational approach to mimic the receptive field properties of retinal ganglion cells [16]. Applying a Gaussian filter to an image creates a smoothed or blurred version of it. The Difference of Gaussians (DoG) output is obtained by subtracting two differently blurred versions of the same image, effectively functioning as a band-pass filter. The most efficient way to calculate the DoG output is by first generating the DoG filter itself and then applying it directly to the image in a single convolution.

The DoG output of the retinal ganglion cells model, incorporating a centre-surround organisation for a 2D image like *I*, is expressed as:

$$\Gamma_{\sigma,s\delta}(x,y) = I \times \frac{1}{2\pi\sigma^2} exp[-(x^2+y^2)/2\sigma^2] - I \times \frac{1}{2\pi s^2\sigma^2} exp[-(x^2+y^2)/2s^2\sigma^2]$$

$$Where \ s = {\sigma_{surround}}/{\sigma_{center}} = {\sigma_s}/{\sigma_c}$$

Here, $x$ and $y$ denote the horizontal and vertical distances from the origin, respectively, while $\sigma$ corresponds to $\sigma_c$, the sigma of the centre Gaussian. The sigma of the surround Gaussian is represented by $\sigma_s = s\sigma$, where *s* is the *Surround ratio*. Therefore, based on the *s* factor, the ratio of the surround Gaussian to the centre Gaussian is defined. This concentric arrangement of the centre and surround Gaussians models the retinal point spread function (PSF) and lateral inhibition (LI) in the retina [29-31]. This structure captures the antagonistic center-surround organization of ganglion cells, which are sensitive to local contrast variations.

The second derivative of the Gaussian, commonly estimated by the difference of two DoGs, is known as the Laplacian of the Gaussian (LoG). Research has demonstrated that for modelling the receptive fields of retinal ganglion cells, DoG [31,32] provides a close approximation to LoG when the dispersion ratio of the centre to surround ($s$) is approximately 1.6 ($\approx \phi$, the Golden Ratio) [25]. As $s$ increases, the area covered by surround suppression broadens while its peak height decreases. In the experimental trials reported here, $s = 1.6$ ($\approx \phi$) is used.

In practice, handling Gaussians with an infinite range is impractical, so the DoG model is constrained within a window where the Gaussian values are minimal beyond its bounds (falling below 5% for the surround Gaussian). To manage computational load, the window size (*windowSize*) is controlled, as larger windows are more computationally intensive. This *windowSize* is determined by the parameter $h$ (*Window ratio*) and $\sigma_c$, as shown below:

$$windowSize = h \times \sigma_c + 1$$

The parameter *h* controls the extent of the center and surround Gaussians included in the filter. In this paper, we standardise by setting $h = 8$

Fig. 2 presents both 2D and 3D visualisations of a sample Difference of Gaussians (DoG) filter used in this study. The parameters for this filter are specified as follows: the standard deviation ($\sigma_c$) of the centre Gaussian is set to 16, while the *Surround ratio* is 1.6, resulting in a standard deviation ($\sigma_s$) of 25.6 for the surround Gaussian. The filter size of 129x129 pixels is determined by the *windowSize* parameter, which is set to 8 in our proposed model, ensuring nearly all the surround Gaussian is included. For computational efficiency, this *windowSize* can be reduced to 6.4, with only 5% of the surround Gaussian omitted. The size of investigated patterns/illusions are 1600 x 899 px.

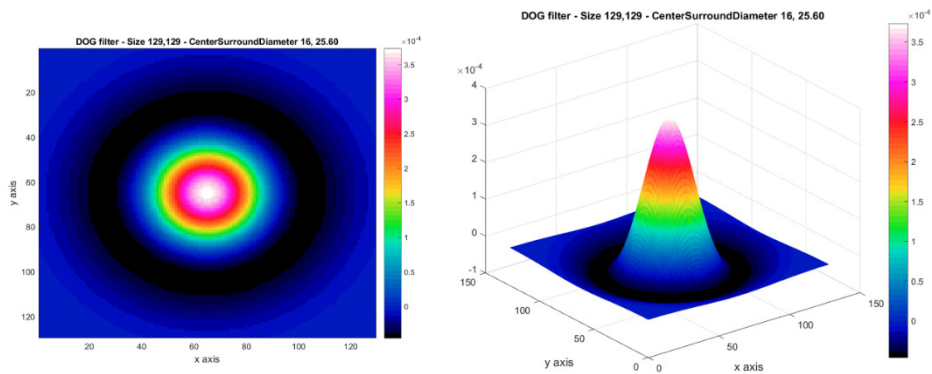

*Fig. 2.* (Left) *2D representation and* (Right) *3D representation of a sample DoG filter. Parameters include* $\sigma_c$ *= 16, for the centre Gaussian, a* Surround ratio *of 1.6, making* $\sigma_s$ *= 25.6. The filter size is 129x129 pixels, based on a* windowSize *parameter of 8.*

### 3.2. Modelling the Expanding Hole Illusion

To model the *Expanding Hole Illusion*, a series of concentric DoG filters were applied to synthetic versions of the illusion pattern to simulate the responses of retinal ganglion cells. The DoG filters were tuned to respond to varying contrast gradients within the pattern, simulating the differential activation of retinal ganglion cells. The parameters of the filters, including $\sigma = \sigma_c$ and $s\sigma = \sigma_s$, were adjusted to mimic the varying sensitivity of retinal cells to different contrast gradients [17,19]. The output of the DoG layers was aggregated to produce a map of contrast responses across the visual field as we refer to it as an "*Edge-Map*".

Figs. 3 and 4 provide visualisations of the *Expanding Hole Illusion* and its contrast-polarity version, using a Difference of Gaussians (DoG) Edge-Map. In Fig. 3, the Edge-Map captures the illusion's expansion effect by displaying changes in the filtering response at progressively increasing scales in the central region. The scale of the centre Gaussian ($\sigma_c$) increases from 4 to 20 in increments of 4, intensifying the DoG filter's sensitivity to contrasts in the central area and simulating the dynamic effect of forward motion. This emphasised response visually reinforces the perceived expansion in the original pattern.

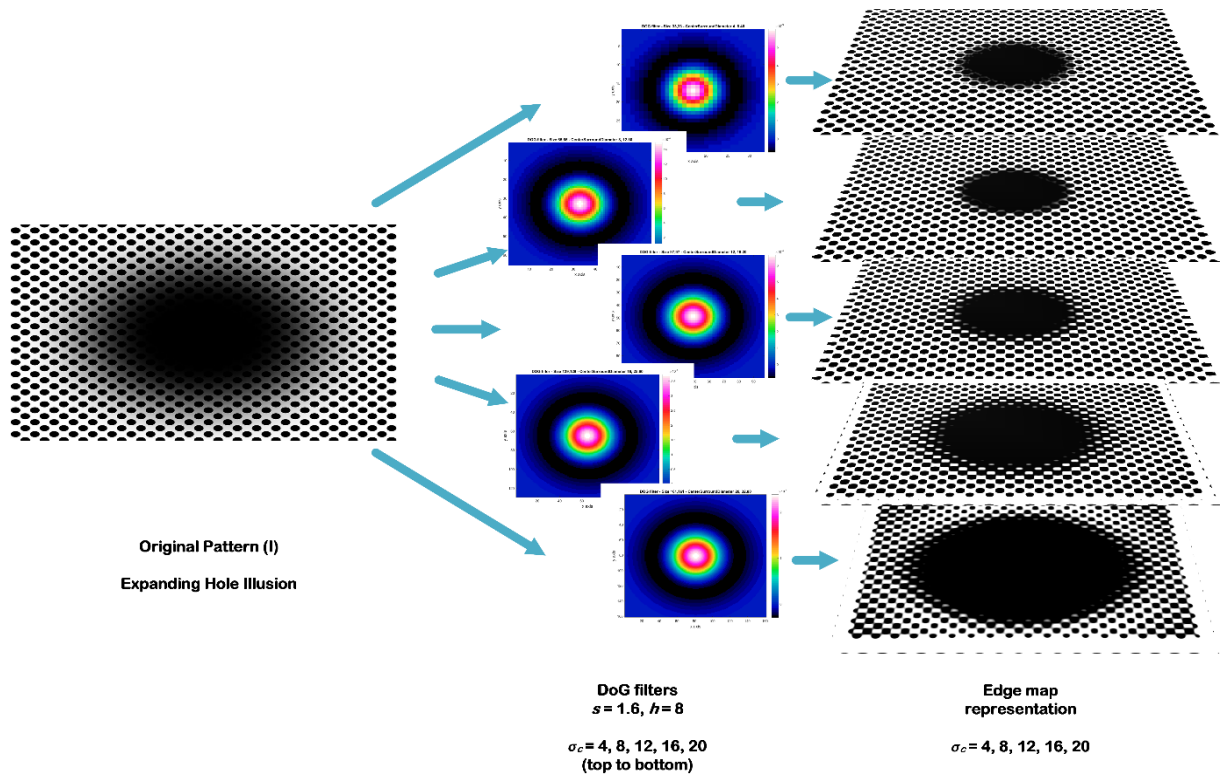

Fig. 3. DoG Edge-Map representation of the Expanding Hole Illusion (Right). This Edge-Map highlights the expansion effect in the central region by showing the filtering response as the scale of σ_c increases from 4 to 20 in steps of 4.

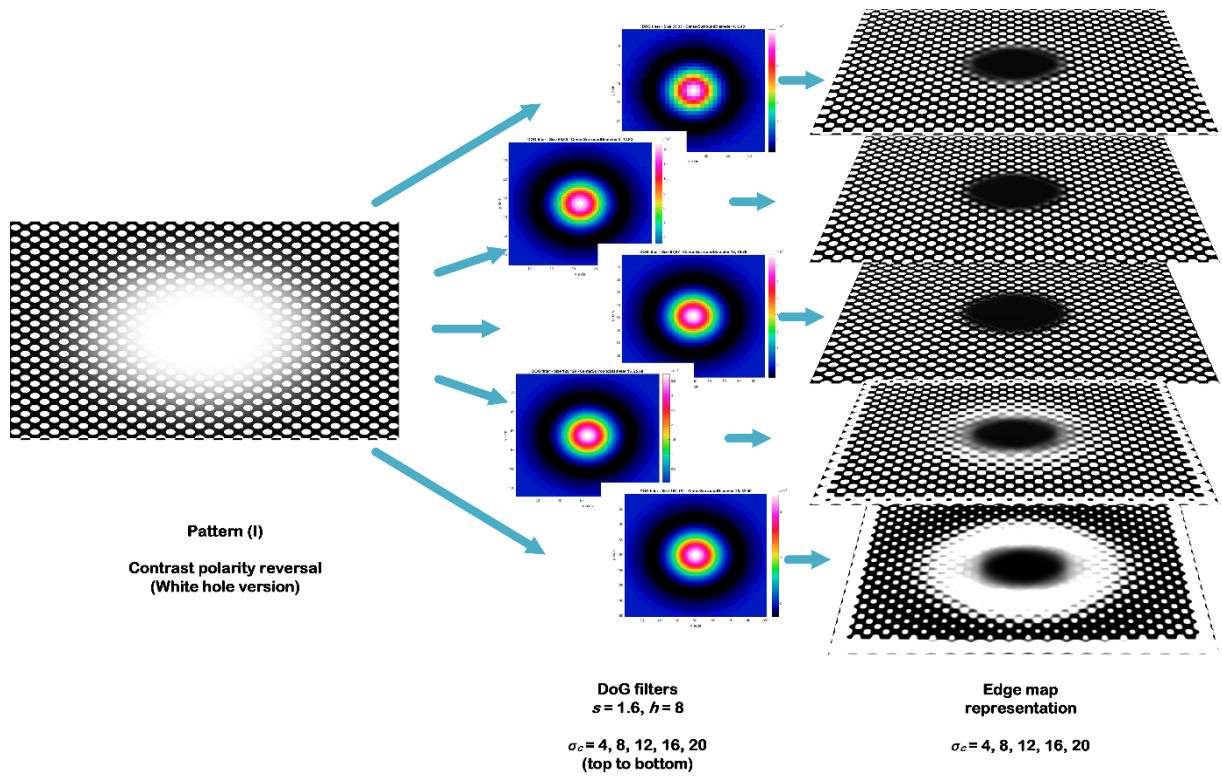

Fig. 4. DoG Edge-Map representation of the contrast polarity reversal – White Hole illusion (Right). This Edge-Map highlights the shrinkage effect in the central region by showing the filtering response as the scale of σ_c increases from 4 to 20 in steps of 4.

In contrast, Fig. 4 shows that, in the contrast polarity version, the central region's expansion is not observed in the Edge-Map as the scale increases. Even at coarser scales ($\sigma_c$ = 16 and 20), the filtered response shows a contraction of the central region, reflecting a perceptual impression of inward movement rather than outward expansion.

### 3.3. Computational Framework

The model was implemented in MATLAB, leveraging custom scripts to process the synthetic pattern and visualise the DoG response maps. The output of the DoG filters was aggregated and compared against the patterns reported by human observers in the psychophysical experiments conducted by Laeng et al. [1].

## 4. Results

### 4.1. Simulated Perception of Motion

The DoG-based receptive field model demonstrated that increased lateral inhibition at higher contrast levels could account for the perceived expansion, aligning with previous findings on motion illusions [6,9,13]. Furthermore, integrating pupil dynamics into the model predictions allowed for an enhanced understanding of the illusion's interaction with physiological processes.

The DoG-based model successfully replicated the perceived expansion effect of the *Expanding Hole Illusion*. The filtering process produced localized increases in neural activity, particularly at the edges of the expanding pattern, where contrast gradients were highest. These findings align with the hypothesis that retinal ganglion cells contribute to the perception of illusory expansion through their contrast-sensitive properties [19]. We further explore the filtering response by examining the LoG responses of the previous Edge-Maps.

Fig. 5 presents the Laplacian of Gaussian (LoG) Edge-Maps, illustrating the *Expanding Hole Illusion* on the left and its contrast polarity version, the "*White Hole*" Illusion, on the right. These LoG Edge-Maps capture the perceptual cues related to the expansion and contraction effects within the illusions, achieved by using different scale settings.

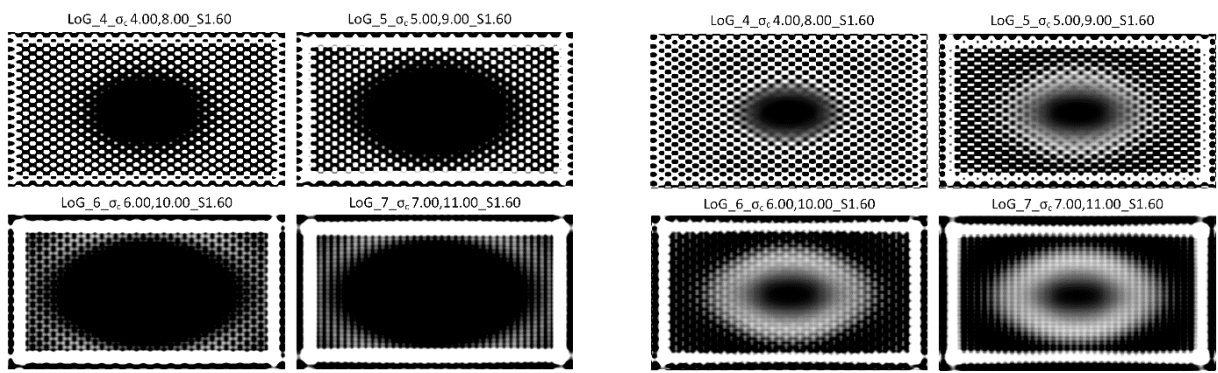

*Fig. 5. Laplacian of Gaussian (LoG) Edge-Maps for the Expanding Hole Illusion* (Left) *and its contrast polarity version, the "White Hole" Illusion* (Right)*. The left panel shows the LoG Edge-Map as the scale increases from the difference between* σ$_c$ *= 4 and* σ$_c$ *= 8 to* σ$_c$ *=7 and* σ$_c$ *=11, highlighting expansion cues in the central region of the pattern. The right panel illustrates the LoG Edge-Map for the White Hole version, where the filtered response highlights the contraction of the central region.*

For the *Expanding Hole Illusion* (left panel), the LoG Edge-Map displays responses at two different scale intervals: $\sigma_c$ = 4 and $\sigma_c$ = 8 to $\sigma_c$ = 7 and $\sigma_c$ =11. As the scale of the Gaussian differences increases, the central region appears to expand, capturing the dynamic illusion of forward movement into a dark tunnel. This effect is emphasised by the increased response at larger scales, which draws attention to the contrast boundaries in the central region and reinforces the visual impression of expansion.

In contrast, the *White Hole* version (right panel) produces a LoG Edge-Map that captures cues for contraction in the central area. This edge map shows that as the scale increases, the central region appears to shrink, suggesting a visual effect of moving towards a bright light. The LoG filter's response in this polarity-reversed version accentuates the inward contrasts rather than outward, visually supporting the perceived contraction in the centre of the pattern.

*4.2. Comparison with Psychophysical Data*

The simulated response maps were compared with the physiological data reported by Laeng et al. [1]. The model outputs closely mirrored the regions of the pattern that triggered significant pupil dilation in human participants. This correspondence suggests that the illusion arises from early retinal processing rather than higher-level visual or cognitive mechanisms.

# 5. Discussion

The findings indicate that the *Expanding Hole Illusion* is a result of complex neural interactions, specifically involving contrast-based lateral inhibition within the visual cortex. The alignment of perceptual expansion strength with physiological pupil responses suggests that the illusion is not purely perceptual but involves coordinated processing across different neural circuits. The illusion's dynamic nature, despite being presented as a static image, highlights the adaptability of the human visual system and the sensitivity of receptive fields to gradient changes. Our model aligns with previous studies on visual motion illusions, extending these insights to explain the *Expanding Hole Illusion*.

Our findings extend previous research on the Rotating Snakes illusion and other dynamic illusions, where differential activation of receptive fields created the perception of motion in a static image [9,13]. Additionally, our results support Laeng's [1] hypothesis that such illusions may activate physiological mechanisms due to the brain's interpretation of the image as a genuine motion stimulus.

*5.1. Implications for Understanding Illusions*

Our findings support the notion that early retinal processing plays a key role in generating the *Expanding Hole Illusion*. The alignment between our model and Laeng's [1] psychophysical data highlights the importance of contrast-sensitive mechanisms in the retina. Specifically, the model suggests that retinal contrast mechanisms may serve as the basis for misinterpreting static patterns as expanding. This reinforces the idea that visual illusions can originate from the inherent properties of early neural processing, rather than solely relying on higher-order cognitive interpretations [7].

The findings support the broader hypothesis that contrast-based interactions in the visual pathway are integral to the interpretation of static patterns as motion stimuli. This ties into the framework of contrast adaptation, previously discussed by Anstis et al. [33], who suggested that the perception of motion in static images can be tied to mechanisms of local luminance adaptation in the retina and cortex.

Here we should note that the *Expanding Hole Illusion* shares essential characteristics with the "*Tunnel Motion*" effect reported in [34]. Both illusions rely on dark gradients and central regions that evoke a perception of depth and forward motion. In the case of the *Expanding Hole Illusion*, observers perceive the central area as expanding, creating a sense of movement inward. This illusion appears to tap into similar low-level visual processing mechanisms as those involved in Laeng and Kitaoka's *Tunnel Motion* effect.

This study extends previous work by proposing that illusions involving apparent expansion could be tied to the way retinal cells encode visual information. Our findings also support the idea that neural processing at the early stages of the visual pathway can contribute to illusory motion experiences.

*5.2. Neural Correlates of Illusory Motion*

Recent neurophysiological studies have identified contrast-sensitive neurons in the retina and primary visual cortex (V1) as key players in motion perception [9,19]. Our computational results lend credence to the hypothesis that similar neural circuits may contribute to illusory motion experiences. By demonstrating that the DoG filter can replicate the *Expanding Hole Illusion*, we extend the understanding of how retinal ganglion cells process complex visual stimuli.

*5.3. Limitations and Future Directions*

While our model provides a plausible explanation for the *Expanding Hole Illusion*, further work is needed to verify these findings through neuroimaging studies and more extensive psychophysical experiments. For instance, recording the activity of retinal cells in response to the illusion could provide direct evidence of contrast processing as the primary mechanism.

Additionally, it would be valuable to investigate the involvement of other visual areas, such as V1 and MT, in modulating the perception of expansion. Future research could also explore how this model generalises to other types of visual illusions involving motion such as *Tunnel Motion* effect [34].

# 6. Conclusion

The *Expanding Hole Illusion* challenges traditional views of motion perception by demonstrating how static images can evoke strong sensations of movement. Through a combination of previously reported psychophysical experiments and our work in bioderived modelling, we have shown that this illusion likely arises from contrast-dependent lateral interactions in early visual areas.

This study presents a computational model based on DoG filtering that provides an explanation for the *Expanding Hole Illusion*. By simulating early retinal processing, the model accounts for the perception of expansion as a byproduct of contrast sensitivity. Our study provides evidence supporting a receptive field-based explanation for the illusion and offers a computational framework to model its underlying neural mechanisms. The close alignment between our results and Laeng's psychophysical findings emphasizes the importance of retinal mechanisms in generating illusory motion perceptions.

Moreover, the illusion's correlation with physiological responses such as pupil dilation underscores its impact on multiple neural and perceptual pathways. Our findings contribute to a growing body of research on dynamic illusions such as "*Tunnel Motion*" effect [34] and offer a new framework for understanding the neural basis of perceptual biases. The insights gained from this model offer a broader understanding of visual processing and highlight the significance of contrast-sensitive neural circuits in creating illusory motion.

# References

1.      Laeng B, Nabil, S., Kitaoka, A. The Eye Pupil Adjusts to Illusorily Expanding Holes. *Front Hum Neurosci*. 2022;16:877249. doi:10.3389/fnhum.2022.877249

2. Eagleman DM. Visual illusions and neurobiology. *Nat Rev Neurosci*. Dec 2001;2(12):920-6. doi:10.1038/35104092
3. Hering E. *Outlines of a theory of the light sense*. Harvard University Press; 1964.
4. Changizi MA, Hsieh A, Nijhawan R, Kanai R, Shimojo S. Perceiving the present and a systematization of illusions. *Cogn Sci*. Apr 5 2008;32(3):459-503. doi:10.1080/03640210802035191
5. Nematzadeh N, Powers DMW. A Predictive Account of Cafe Wall Illusions Using a Quantitative Model. *arXiv*. 2017;doi:https://doi.org/10.48550/arXiv.1705.06846
6. Nematzadeh N, Powers DMW, Lewis T. Informing Computer Vision with Optical Illusions. *arXiv*. 2019;doi:https://doi.org/10.48550/arXiv.1902.02922
7. Gregory RL. *Eye and brain: the psychology of seeing* (5th ed) Oxford University Press. 1997.
8. Hubel DH, Wiesel TN. Receptive fields, binocular interaction and functional architecture in the cat's visual cortex. *J Physiol*. Jan 1962;160(1):106-54. doi:10.1113/jphysiol.1962.sp006837
9. Conway BR, Kitaoka A, Yazdanbakhsh A, Pack CC, Livingstone MS. Neural basis for a powerful static motion illusion. *J Neurosci*. Jun 8 2005;25(23):5651-6. doi:10.1523/JNEUROSCI.1084-05.2005
10. Faubert J, Herbert AM. The peripheral drift illusion: a motion illusion in the visual periphery. *Perception*. 1999;28(5):617-21. doi:10.1068/p2825
11. Whitney D, Cavanagh P. Motion distorts visual space: shifting the perceived position of remote stationary objects. *Nat Neurosci*. Sep 2000;3(9):954-9. doi:10.1038/78878
12. De Valois RL, De Valois KK. Spatial vision. *Oxford psychology series, No 14 Spatial vision New York, NY, US: Oxford University Press*. 1988;
13. Kitaoka A. The Fraser-Wilcox illusion group: its phenomena and models. . *IEICE Technical Report; IEICE Tech Rep,*. 2012;112(168), pp.57-60
14. Larsen A, Madsen KH, Lund TE, Bundesen C. Images of illusory motion in primary visual cortex. *J Cogn Neurosci*. Jul 2006;18(7):1174-80. doi:10.1162/jocn.2006.18.7.1174
15. Kitaoka A. Color-Dependent Motion Illusions in Stationary Images and Their Phenomenal Dimorphism. *Perception,* . 2014;43(9), 914-925.doi: https://doi.org/10.1068/p7706
16. Rodieck RW. Quantitative analysis of cat retinal ganglion cell response to visual stimuli. *Vision Research (Oxford)*. 1965; 5(12), 583–601. doi:https://doi.org/10.1016/0042-6989(65)90033-7
17. Nematzadeh N. *A Neurophysiology Model that Makes Quantifiable Predictions of Geometric Visual Illusions (Doctoral dissertation, Ph. D. Dissertation. Flinders University of South Australia)*. 2018.
18. Nematzadeh N, Lewis T, Powers DMW. Bioplausible multiscale filtering in retinal to cortical processing as a model of computer vision. . *ICAART2015-International Conference on Agents and Artificial Intelligence SCITEPRESS*. 2015;
19. Carandini M, Demb JB, Mante V, et al. Do we know what the early visual system does? *J Neurosci*. Nov 16 2005;25(46):10577-97. doi:10.1523/JNEUROSCI.3726-05.2005
20. Choi H, Kim J, Lee H. Modelling motion perception using differential motion detectors. *Journal of Vision,* . 2019;19(10), 1-14.
21. Pinna B, Brelstaff GJ. A new visual illusion of relative motion. *Vision Res*. 2000;40(16):2091-6. doi:10.1016/s0042-6989(00)00072-9
22. Field GD, Chichilnisky EJ. Information processing in the primate retina: circuitry and coding. *Annu Rev Neurosci*. 2007;30:1-30. doi:10.1146/annurev.neuro.30.051606.094252
23. Gollisch T, Meister M. Eye smarter than scientists believed: neural computations in circuits of the retina. *Neuron*. 2010;65.2 150-164.
24. Marr D. Vision. A computational investigation into the human representation and processing of visual information *WH San Francisco: Freeman and Company, 1(2)*. 1982;WH San Francisco: Freeman and Company.
25. Marr D, Hildreth E. Theory of edge detection. *Proc R Soc Lond B Biol Sci*. Feb 29 1980;207(1167):187-217.
26. Cook PB, McReynolds JS. Lateral inhibition in the inner retina is important for spatial tuning of ganglion cells. *Nature neuroscience*. 1998;1(8): 714-719.

27. Ratliff F, Knight BW, Graham N. On Tuning and Amplification by Lateral Inhibition. *Proceedings of the National Academy of Sciences*. 1969;62(3):733-740. doi:10.1073/pnas.62.3.733
28. Huang JY, Protti DA. The impact of inhibitory mechanisms in the inner retina on spatial tuning of RGCs. *Scientific reports*. 2016;6:21966. doi:10.1038/srep21966
29. Enroth-Cugell C, Robson JG. The contrast sensitivity of retinal ganglion cells of the cat. *J Physiol*. Dec 1966;187(3):517-52.
30. Rodieck RW, Stone J. Analysis of Receptive Fields of Cat Retinal Ganlion Cells. *Journal of Neurophysiology, 28(5), 833-849*. 1965;
31. Linsenmeier RA, Frishman LJ, Jakiela HG, Enroth-Cugell C. Receptive field properties of x and y cells in the cat retina derived from contrast sensitivity measurements. *Vision Res*. 1982;22(9):1173-83.
32. Robson JG. Frequency domain visual processing. *Physical and biological processing of images, Springer Berlin Heidelberg*. 1983:73-87.
33. Anstis S, Rogers B. Illusory Motion: A Consequence of Adaptation to Luminance *Nature,* . 2000;408(6815), 184-188.doi: https://doi.org/10.1038/35041588
34. Laeng B, Nabil S, Kitaoka A. Tunnel motion: Pupil dilations to optic flow within illusory dark holes. *Perception*. Oct 2024;53(10):730-745. doi:10.1177/03010066241270493